\documentclass[12pt]{article}

\usepackage{amsfonts}
\usepackage{amssymb}
\usepackage{amstext}
\usepackage{amsmath}
\usepackage{amsthm}
\usepackage{bm}
\usepackage{caption}
\usepackage{graphicx}
\usepackage{slashed}
\usepackage[linkcolor=blue]{hyperref}

\begin{document}

%%%%%%%%%%%%%%%%%%%%%%%% (AMS)LATEX MACROS %%%%%%%%%%%%%%%%%%%%%%%%%

\def\theequation{\thesection.\arabic{equation}}
\def\be{\begin{equation}}
\def\ee{\end{equation}}
\def\ba{\begin{eqnarray}}
\def\ea{\end{eqnarray}}
\def\lb{\label}
\def\nn{\nonumber}

\def\a{\alpha}
\def\b{\beta}
\def\g{\gamma}
\def\d{\delta}
\def\i{\eta}
\def\e{\varepsilon}
\def\l{\lambda}
\def\r{\rho}
\def\s{\sigma}
\def\t{\tau}
\def\o{\omega}
\def\v{\varphi}
\def\x{\xi}

\def\D{\Delta}
\def\G{\Gamma}
\def\O{\Omega}
\def\L{\Lambda}

\def\fU{\mathfrak A}
\def\bo{\mathfrak b}
\def\fp{\mathfrak p}
\def\hp{\hat p}

\def\E{{\cal E}}
\def\Vp{{\cal V}_p}
\def\Hp{{\cal H}_p}

\def\bq{\overline{q}}
\def\bM{\bar M}
\def\bz{\bar z}

\def\bU{\overline{U}_q}
\def\bbU{{\overline {\overline U}}_q}
\def\tU{{\widetilde{U}}_q}

\def\bD{\overline {\cal D}}

\def\fl{\mathfrak l}
\def\fr{\mathfrak r}

\def\bd{\bf d}
\def\sJ{^*\!\!{\cal J}}

\def\ux{\underline x}
\def\up{\underline p}
\def\uq{\underline q}
\def\uz{\underline z}
\def\pz{\Pi_{23}\;\uz}

\def\id{\mbox{\em 1\hspace{-3.4pt}I}}
\def\idd{\scriptsize{\mathit 1 \hspace{-4.2pt}I}}
\newcommand{\ID}[2]{\id^{| #1 {\cal i}}_{\;\;\; {\cal h} #2 |}}

\def\p{\hat p}

\def\Z{\mathbb Z}
\def\R{\mathbb R}
\def\C{\mathbb C}
\def\F{\mathbb F}
\def\H{\mathbb H}

\def\1{1\!\!{\rm I}}

\def\eod{\phantom{a}\hfill \rule{2.5mm}{2.5mm}}

\def\hR{\hat{R}}
\def\Rp{\hat{R}(p)}
\def\Rcp{\hat{R}^c(p)}

\def\hp{\hat p}
\def\fp{\mathfrak p}
\def\sp{\slash\!\!\!{p}}
\def\sfp{\slash\!\!\!{\mathfrak p}}
\def\sk{\slash\!\!\!{k}}
\def\sq{\slash\!\!\!{q}}
\def\sd{\slashed\partial}
\def\sA{\slash\!\!\!\! {A}}
\def\sD{\slash\!\!\!\! {D}}

\def\subbbc{{\rm C}\kern-3.3pt\hbox{\vrule height4.8pt width0.4pt}\,}
\def\qd{\stackrel{.}{q}}
\def\pl{\partial}
\def\vac{\mid 0 \rangle}
\def\lvac{\langle 0 \!\mid}

\def\Ba{\bar a}
\def\Bp{\bar p}

\def\Fd{{\cal F}^{diag}}
\def\Fp{{\cal F}'}
\def\bm{\mathbf m}
\def\bp{\mathbf p}

%%%%%%%%%%%%%%%%%%%%%%%%%%%%%%%%%%%%%%%%%%%%%%%%%%%%%%%%%%%

\begin{titlepage}

%\begin{flushright}DRAFT, \ {\today}\end{flushright}\vskip 2cm

\begin{center}

{\Large{\bf{Neutrino, parity  violaton, V-A: \\ a historical survey}}}

\vspace{8mm}

{\large{\bf{Ludmil Hadjiivanov}}}

\vspace{4mm}

Institute for Nuclear Research and Nuclear Energy\\
Bulgarian Academy of Sciences\\
Tsarigradsko Chaussee 72, BG-1784 Sofia, Bulgaria\\
e-mail: lhadji@inrne.bas.bg

\end{center}

\vspace{5mm}

\begin{abstract}

This is a concise story of the rise of the four fermion theory of the universal weak interaction
and its experimental confirmation, with a special emphasis on the problems related to parity violation.

\end{abstract}

\end{titlepage}

\newpage

\tableofcontents

\newpage

\section*{Foreword}
\addcontentsline{toc}{section}{Foreword}

This article emerged initially as a part of a bigger project (joint with Ivan Todorov, still under construction) on the foundations of the Standard Model of particle physics and the algebraic structures behind its symmetries, based in turn on a lecture course presented by Ivan Todorov at the Physics Department of the University of Sofia in the fall of 2015.

The first controversies found in beta decay properties were successfully resolved in the early 30-ies by Pauli and Fermi. The bold prediction of Lee and Yang from 1956 that parity might not be conserved in weak interactions has been confirmed in the beginning of 1957 by the teams of Mme Wu and Lederman and nailed by the end of the same year by Goldhaber, Grodzins and Sunyar who showed in a fine experiment that the neutrino was indeed left-handed. The present survey covers the development in this field till 1958 when the universal chiral invariant (V-A) four fermion interaction proposed by Sudarshan and Marshak, Feynman and Gell-Mann was widely accepted.

Certainly, these glorious days for particle physics have attracted attention from many researchers and historians. The author has learned a lot, in particular, from the beautiful narrative of Jagdish Mehra \cite{Me} as well from \cite{Fr79, M, Fo, J00, L09}, among others. The written memories of some of the main participants in these events like Yang (e.g. in \cite{Y82}), Feynman \cite{FL}, Sudarshan and Marshak \cite{SM3, SM4} and also of S. Weinberg \cite{W09} and S. Glashow \cite{G09} provide both first hand expert evidence and evaluation of the discoveries from the position of the elapsed time.

For everyone honestly interested in understanding how this precious cornerstone of the present day Standard Model was put in its place, finding and reading the sources is a thrilling experience. It is astonishing, for example, to learn that parity violation could have been discovered already in 1928-1930 by Cox and Chase, see Remark 2 below, or to follow Feynman on his way toward V-A, and it is very impressive to see how theoreticians and experimenters from particle and  nuclear physics collaborated to reach the solutions of some of the most difficult puzzles provided by nature.

\newpage

\section{The beta decay and the neutrino}
\setcounter{equation}{0}
\renewcommand\theequation{\thesection.\arabic{equation}}

Soon after radioactivity was discovered by Antoine Henri Becquerel in 1896 and studied further by Pierre and Marie Curie,
Ernest Rutherford, Becquerel and Paul Villard identified the three types of observed emissions, $\alpha\,, \beta$ and $\gamma\,,$
as ions of helium, J.J. Thompson's electrons and high energy electromagnetic radiation, respectively.
The nuclear beta decay, in particular, is a manifestation of the {\em weak} interactions whose understanding required
several unexpected, quite radical changes in our assumptions about the basic laws of nature.

In 1913 Niels Bohr suggested correctly that it was the nucleus discovered by E. Rutherford in 1911 that was "the seat of
the expulsion of high speed beta particles" (rather than the electronic distribution round it)\footnote{N. Bohr,
On the constitution of atoms and molecules, Part II. Systems Containing Only a Single Nucleus,
{\it Phil. Mag.} {\bf 26} (1913) 476-502.}. It was found that the kinetic energy of the emitted electrons varied from almost
zero to ultra relativistic levels and moreover (James Chadwick 1914), had a continuous spectrum\footnote{J. Chadwick,
Intensit\"atsverteilung im magnetischen Spektrum von $\b$-Strahlen von Radium B+C, {\it Verhandl. Dtsch. Phys. Ges.}
%Verhandlungen der Deutschen Physikalischen Gesellschaft (in German).
{\bf 16} (1914) 383-391.}. In 1922 Lise Meitner noticed that, if the energy levels of an atomic nucleus are quantized,
such a continuous distribution would pose a real puzzle\footnote{L. Meitner,
\"Uber die Entstehung der $\b$-Strahl-Spektren radioaktiver Substanzen, Z. Physik, {\bf 9}:1 (1922) 131-144.}.
After the discovery of the spin, another confusion arose from examples of beta decays in which initial and final nuclei
both had integer angular momentum, thus contradicting the statistics rules provided that the electron, of spin $1/2\,,$ was
the only emitted particle detected in experiments. To resolve all this mess, ideas as reckless as that of energy non-conservation
were proposed (by N. Bohr \cite{B30}, and later by W. Heisenberg \cite{H32}!\,\footnote{Note that the BKS theory of radiation, see N. Bohr, H.A. Kramers, J.C. Slater, The quantum theory of radiation, {\it Phil. Mag.} {\bf 47} (1924) 785-802 (\"Uber die Quantentheorie der Strahlung, {\it Z. Physik}, {\bf 24} (1924) 69-87, in German) reduced conservation of energy to
a statistical law, but these were still the times of the "old quantum theory".}). One should have in mind that by 1930,
the only known "elementary particles" were the electron and the proton (the nucleus of hydrogen, E. Rutherford 1920),
and that Dirac's theory had not been yet confirmed and widely accepted.

A possible solution of the puzzle, at once explaining the continuous energy spectrum and correcting the energy balance and the seemingly wrong statistics, was submitted in the famous one-page "Open letter to the group of radioactive people", sent by Wolfgang Pauli to the participants at the meeting in T\"ubingen on 4 December 1930. In it Pauli proposed that "{\it in the nuclei} there could exist electrically neutral particles, {\it which I will call neutrons}, that have spin 1/2 and obey the exclusion principle and that further differ from light quanta in that they do not travel with the velocity of light." \cite{P30}. Pauli thus assumed that in beta decay, in addition to the electron, this yet unknown, light neutral particle of half integer spin was emitted so that the sum of energies of this particle and the electron was constant (equal to the upper limit of the $\b\,$ energy spectrum). Experimental evidence required that the new particle should have a very large ability to get through matter. In June 1931 Pauli announced his idea at the Pasadena American Physical Society meeting where it was met with scepticism, and discussed it later in October at the Rome Nuclear Physics conference with Enrico Fermi who, by contrast, was immediately attracted by it.

In December that same year Carl Anderson discovered the positron, and in February 1932 J. Chadwick announced the possible existence
of "particles of mass $1\,$ and charge $0\,,$ or neutrons" \cite{Ch32}. Neutrons were actually observed
(but erroneously interpreted as high energy $\g$ quanta) earlier, when a radiation emitted from beryllium bombarded by
$\a$-particles of polonium was registered by Walther Bothe and Herbert Becker in Berlin\footnote{W. Bothe,
H. Becker, K\"unstliche Erregungen von Kern $\g$-Strahlen, {\it Z. Phys.} {\bf 66} (1930) 289-306.}
and then by Ir\`ene Curie and Fr\'ed\'eric Joliot in Paris\footnote{
I. Curie, F. Joliot, \'Emission de protons de grande vitesse par les substances hydrog\'en\'ees sous l'influence des rayons $\a$
tr\`es p\'en\'etrants,
%The emission of high energy photons from hydrogenous substances irradiated with very penetrating alpha rays,
{\it C. R. Acad. Sci. Paris} {\bf 194} (1932) 273-275.}. Similar studies
%\footnote{M. Blau, E. Kara-Mihailova, \"Uber die durchdringende Strahlung des Poloniums, {\it Sitzungsber. Akad. Wiss. Wien IIa} {\bf 140}:8 (1931) 615-622.}
were carried out simultaneously at the Institute of Radium Studies in Vienna by a group including, in particular, the Austrian Marietta Blau and Elisaveta (Elisabeth) Karamichailova, born in 1897 in Vienna
in the family of the Bulgarian Ivan Mikhaylov and the English Mary Slade \footnote{See e.g.
M. Rentetzi, Gender, Politics, and Radioactivity Research in Vienna, 1910-1938, PhD Thesis,
%defended on March 25, 2003 at Virginia Tech, Blacksburg, Virginia,
Virginia Tech, Blacksburg (2003), available at \href{https://vtechworks.lib.vt.edu/handle/10919/27084}{\footnotesize{\tt https://vtechworks.lib.vt.edu/handle/10919/27084}}.}.
For a recent account of the contribution of the young Italian genius Ettore Majorana to both the theory and the interpretation
of these experiments, see \cite{Re} and references therein.

In his contribution to the $7$th Solvay Conference in Brussels (October 1933) Pauli already used the name "neutrino"
%%proposed by Fermi 
to distinguish the light particle from the heavy neutron. The decisive step to a theory of the beta decay has been taken soon after that by Enrico Fermi himself \cite{F33} who proposed his famous {\it four fermion} interaction Hamiltonian
\be
\frac{G_F}{\sqrt{2}}\, (\widetilde p (x) \g_\mu n (x) )\,(\widetilde e (x) \g^\mu \nu (x)) + h.c. \ ,
\lb{Fermi}
\ee
$p\,, n\,, e\,$ and $\nu\,$ standing for the proton, neutron, electron and neutrino, respectively. The structure of (\ref{Fermi}) imitated the newly discovered second quantization approach to the theory of photon
radiation: a vector coupling involving a (charged) current of heavy particles, analogous to the EM one (with {\it no derivatives} of any of the fields), creating an electron-neutrino pair (instead of a photon) in such a way that charge conservation is guaranteed. In contrast to Pauli who assumed that the light particles are also residing in the nucleus, Fermi followed Heisenberg
\cite{H32} in admitting that all nuclei only consist of protons and neutrons (considered as two different quantum states of a heavy particle), and further supposed that an electron and a neutrino are created in every transition from a neutron into a proton -- or destroyed, in the opposite process.

\medskip

\noindent
{\bf Remark 1. The story of the name "neutrino"~} The name "neutrino" with its Italian flavor is usually attributed to E. Fermi. In fact it has been  pronounced for the first time by Edoardo Amaldi (1908 -- 1989)\footnote{The author thanks Serguey Petcov for providing evidence on this point.}, see the remark under [277] in the list of references of \cite{A84}: "{\it The name 'neutrino' (a funny and grammatically incorrect contraction of 'little neutron' in Italian: neutronino) entered the international terminology through Fermi, who started to use it sometime between the conference in Paris in July 1932 and the Solvay conference in October 1933 where Pauli used it. The word came out in a humorous conversation at the Istituto di Via Panisperna. Fermi, Amaldi and a few others were present and Fermi was explaining Pauli's hypothesis about his 'light neutron'. For distinguishing this particle from the Chadwick neutron Amaldi jokingly used this funny name, -- says Occhialini, who recalls of having shortly later told around this little story in Cambridge}."

\medskip

It is worth mentioning that in \cite{F33} Fermi also proposed the direct method of measuring the neutrino mass by examining the energy spectrum of electrons emitted in $\b\,$ decay near its kinematic end point which was subsequently used in a number of experiments.\footnote{The last experiment of this type, the Karlsruhe Tritium Neutrino Experiment KATRIN \cite{K2018} (with the unprecedented sensitivity of 0.2 eV) has been inaugurated on June 11, 2018.} 
 
\smallskip

The vector interaction implies that the nuclear angular momentum doesn't change and hence, the spins of the emitted leptons are antiparallel. Introducing an {\it axial vector} interaction term, G. Gamow and E. Teller \cite{GT36} incorporated also observed decays with $\Delta J = 1\,$ change in the total angular momentum of the nucleus.
%%and respectively, of parallel spins of the $\b$-particle and the               %%(anti)neutrino.
In this dual form the model was able to serve in the next %about twenty
years as {\it the} theory of weak interactions.

\smallskip

Meanwhile, in 1935 Hideki Yukawa \cite {Y35} predicted the existence, and also the mass, of the "meson" carrying the force needed to hold the particles forming the atomic nuclei together. In the same next year 1936 when Carl Anderson received the Nobel prize for the positron, during his studies of cosmic rays by cloud chambers he discovered, together with Seth Neddermeyer, a new particle of nearly the "proper" meson mass. This gave rise to a great confusion since it  was actually the muon $\mu\,,$ a heavier analog of the electron. The correct Yukawa particle, the pion $\pi\,,$ was identified after World War II, in 1947, by the Bristol group led by C.F. Powell in traces of cosmic rays left on photographic emulsions exposed for several months on mountain tops
in the Alps and in the Andes.
% where the cosmic rays are more intense and
%C. Lattes, H. Muirhead, G. Occhialini and C. Powell 
%from the University of Bristol in England
%who investigated traces of cosmic rays left on photographic films in the Andes.
Yukawa and Powell received the Nobel Prizes in Physics in 1949 and 1950, respectively.

By the end of the 1940's measurements of the energy spectrum of electrons produced in muon decay favoured the assumption (suggested by B. Pontecorvo\footnote{B. Pontecorvo was the first to understand that the muon was a "heavy electron", see \cite{P47} where he also formulated the idea of $\mu - e\,$ universality of the Fermi interaction (cf. Section 4 of \cite{P94}).}, O. Klein, G. Puppi, L. Michel) that the latter, like the beta decay,
is actually a three-particle process. Moreover, the $\mu\,$ lifetime was found to be "consistent with the hypothesis that $\mu$-decay and $\b$-decay are separate instances of the same phenomenon, {\em with the same coupling
constant}~" (see \cite{TW49} and references therein). These developments brought to life the notion of "universal Fermi interaction" (UFI). 

In 1949 T.D. Lee, M. Rosenbluth and C.N. Yang \cite{LRY49} noted that, in order to explain why three independent experiments\footnote{The mentioned three independent experiments were the $\b$-decays of nucleons and muons, and the capture of $\mu^-$ by (a proton in) nuclei.}
%(the $\b$-decays of nucleons and muons, as well as the interactions of 
%$\mu^-\,$ with protons)
lead to coupling constants in Fermi type interactions of the same order of magnitude, it would be reasonable to assume that the latter are "transmitted through an intermediate field with respect to which all particles have the same `charge'. The `quanta' of such a field would have a very short lifetime and would have escaped detection." (their idea is known under the name of "Intermediate Vector Boson", or IVB hypothesis). In such a case the original Fermi form of the interaction Lagrangian would be obtained in the limit of infinite mass of the (charged) intermediate field $W_\mu$ \footnote{More precisely, in the IVB interaction Hamiltonian the point Fermi current-current interaction (\ref{Fermi}) is replaced by current-IVB couplings with a common {\em dimensionless} coupling constant $g\,$ obeying $\frac{g^2}{M_W^2} = \frac{G_F}{\sqrt{2}}\,$ where $M_W\,$ is the IVB mass.}.
%\be\frac{g}{2\sqrt{2}}\, \left[ (\bar p (x) \g_\mu n (x) ) + (\bar e (x) \g_\mu %\nu (x)) \right] W^\mu (x) + h.c.\lb{LRY}\ee and $W_\mu (x)\,$ \footnote{}
(The actual lifetime of the weak bosons $W^\pm$ and $Z\,$ discovered at CERN in 1983 is $\sim\, 3. 10^{-25}$ s. The mass of $W^\pm\,$ is $\sim 80$ GeV, and that of $Z\, \sim 91$ GeV.)

In 1947 two new, heavier mesons produced by cosmic rays in a cloud chamber have been observed in Manchester
by G.D. Rochester and C.C. Butler. Both having nearly half of the proton mass, one of these decayed into two, and
the other -- into three pions. The number of newly discovered particles increased quickly in the following years.
The observed selection rules required additional quantum numbers. For example, the notion of {\em strangeness} $S\,$
%(cf. Section 1.2)
was introduced to explain the enormous disproportion (of 13-15 orders in time) between the fast production
(in pairs, in pion-nucleon collisions) and the slow decay of some of them. It was assumed that $S\,$ is conserved in
strong and electromagnetic interactions but could be violated in the weak decays of "strange" particles.
Another, quite unexpected peculiarity of the weak interactions was however almost impossible to grasp.

\section{The notion of parity and the $\theta-\tau$ puzzle}

Despite the increasing precision of measurements, it turned out that two species of mesons\footnote{Actually,
of the same type as those observed by Rochester and Butler in 1947!} share the same masses and lifetimes.
The problem dubbed in the middle of the fifties "the $\theta - \tau\,$ puzzle" arose from the {\em intrinsic parity} $P\,$
mismatch in decays of the type
\be
\theta^+ \ \to\ \pi^+ + \pi^0\ ,\qquad \tau^+ \ \to\ \pi^+ + \pi^+ + \pi^-\ ,
\lb{theta-tau}
\ee
respectively. Anticipating the breakthrough that followed, we may reveal that eventually $\theta^+\,$ and $\tau^+\,$ were
identified with each other and called (in this case, positively charged) $K$-mesons, or kaons.

\smallskip

Generally speaking, (space) parity could be described as the relation between an object and its mirror reflection.
(This is seemingly different from the definition based on reflection in {\em all three} space coordinates,
\be
I_s\, (x^0, {\mathbf x}) = (x^0, - {\mathbf x})\ ,
\lb{spacepar}
\ee
but reduces to it by rotating the mirror at angle $\pi\,.$)
Developments in optics, crystallography, and chemistry led by the beginning of the 19th century to the discovery that
certain crystals or liquids were optically active, rotating the polarization plane of linearly polarized light in a
specific way, clockwise (in ($+$)-isomers, in modern notation) or counter-clockwise (in ($-$)-isomers)\footnote{M. Arago (1811),
J.-B. Biot (1812-1818); see J.F.W. Herschel's \cite{H1820}.
%On the rotation impressed by plates of rock crystal on the planes of polarization of the rays of light,
%as connected with certain peculiarities in its crystallization,
%{\it Transactions of the Cambridge Philosophical Society} {\bf 1} (1820) 43-51.
}. In 1848 Louis Pasteur sorted manually
two different, non-superimposable mirror-image crystal types of the sodium ammonium salt of tartaric acid produced
by chemical synthesis and found that in solution their optical rotations were equal in magnitude but opposite in direction.
(This fact came as a surprise since natural tartaric acid extracted from wine lees only contained the ($+$)-isomer.)
Pasteur correctly attributed these macroscopic properties to the existence of two mirror-image molecule types of tartaric acid,
using the term "dissymmetry" to express the fact that they could not be transformed one into another in a continuous way.
The present day notion of {\em chirality}, first used by Lord Kelvin in 1893 (see \cite{Th1894}), originates from the Greek word
$\chi \varepsilon \iota \rho$ for "hand"; indeed, left or right {\it handedness} is a common manifestation of this property.

\smallskip

The first application of the notion of parity to quantum systems was made in 1924 by the German-born American physicist
Otto Laporte, then doctoral student of the great Arnold Sommerfeld at LMU Munich \cite{L24}.
His studies of iron spectra led to the discovery that atomic states fell into two separate groups so that,
after a single photon emission or absorption, the initial and the final states always belonged to {\em different} groups.

Anticipating later developments, this finding can be interpreted in the following way.
As the quantum mechanical reflection operator $P\,$ generates a multiplicative $\Z_2\,,$ it has eigenvalues $\pm 1\,.$
Grouping the atomic orbitals into subsets of even ($P=1$) and odd ($P = -1$) parity and assigning negative parity
to the photon, Laporte's rule is equivalent to {\em parity conservation} in single photon atomic transitions.
A complete theoretical understanding has been obtained soon after that by E.P. Wigner \cite{W27}.

The following excerpt from C.N. Yang's 1957 Nobel lecture sheds light on the next about 30 years:
{\em "In 1927 E. Wigner took the critical and profound step to prove that the empirical rule of Laporte is a consequence of the
reflection invariance, or right-left symmetry, of the electromagnetic forces in the atom. This fundamental
idea was rapidly absorbed into the language of physics. Since right-left symmetry was unquestioned also
in other interactions, the idea was further taken over into new domains as the subject matter of physics extended
into nuclear reactions, $\b$-decay, meson interactions, and strange-particle physics. One became accustomed to the
idea of nuclear parities as well as atomic parities, and one discusses and measures the intrinsic parities of the mesons.
Throughout these developments the concept of parity and the law of parity conservation proved to be extremely fruitful,
and the success had in turn been taken as a support for the validity of right-left symmetry."} \footnote{As put by
A. Franklin \cite{Fr79} (citing H. Frauenfelder and E.M. Henley's "Nuclear and Particle Physics" (1975)),
the concept of parity conservation has become a sacred cow -- in a sense, literally: "A very amusing early reference
to this occurs in a paper by P. Jordan and R. de L. Kronig, [Movements of the lower jaw of cattle during mastication,]
{\em Nature} {\bf 120} (1927) 807. In this paper Jordan and Kronig note that the chewing motion of cows is not straight up
and down, but is rather either a left-circular or a right-circular motion. They report on a survey of cows in Si{\ae}lland,
Denmark, and observe that 55\% are right-circular and 45\% left-circular, a ratio they regard as consistent with unity."
The original conclusion of Jordan and Kronig was actually the following: "As one sees, the ratio of the two kinds is
approximately unity. The number of observations was, however, scarcely sufficient to make sure if the deviation from
unity is real. Naturally these determinations allow no generalisation with regard to cows of different nationality."}

This common belief also leaned on the observation (made by J. Schwinger, G. L\"uders, W. Pauli, ...) that the
{\em combined} $CPT$ symmetry, i.e. the conservation of the product of $P$ with the operators of charge conjugation
(particle-antiparticle exchange) $C$ and time reversal $T$, follows from the basic principles of QFT. (In Wightman's
axiomatic setting $CPT$ conservation is a theorem following from locality, Lorentz invariance and energy positivity,
see \cite{SW, BLT}; its rigorous proof was first given by R. Jost in 1957.)

In 1953, R. Dalitz and E. Fabri noticed that one can obtain information about the spins and the parities of the
$\theta\,$ and $\tau\,$ particles (\ref{theta-tau}), as the intrinsic parity of the pions has been already determined
as odd, i.e. $-1\,.$ Indeed, if one neglects the relative motion of the $\pi$-mesons, parity conservation would
imply that the parity of $\theta\,$ is $(-1)^2 = 1\,$ and that of $\tau\,,$ $(-1)^3 = -1\,.$

To quote C.N. Yang again, {\it "By the spring of 1956 the accumulated experimental data seemed to unambiguously indicate,
along the lines of reasoning discussed above, that $\theta\,$ and $\tau\,$ do not have the same parity,
and consequently are not the same particle. ... the inference would certainly have been regarded as conclusive, and
in fact more well-founded than many inferences in physics, had it not been for the anomaly of mass and lifetime degeneracies."}
(The masses were found to be equal within a fraction of a percent.) So some thirty years after the missing neutrino energy,
the theory of weak interactions met another challenge of a similar magnitude in the $\theta - \tau\,$ puzzle.

\section{Parity violation}

As a way out T.D. Lee and C.N. Yang \cite{LY2} suggested that strange particles -- like $\theta\,$ and $\tau\,,$ in particular --
appear in doublets of opposite parity.\footnote{If correct, such an idea (named "parity doubling") would provide a solution
of the equal mass problem. Unfortunately, it neither worked for other (pairs of) particles nor did it explain the equal lifetimes.}
A few days after their work has been published, Yang presented it at the Sixth Annual Conference on High Energy Nuclear Physics
held at the University of Rochester on April 3-7, 1956. In the discussion that followed Richard Feynman asked the bold question
what would be the consequences if the parity rule was wrong. (Feynman actually posed it on behalf of Martin Block, a fellow
experimentalist and his roommate at the conference, who was afraid that the audience wouldn't listen to him.)
Although Yang answered that he and Lee had considered the idea without reaching any conclusion, they both apparently felt
"the beginnings of doubt" \cite{M}.

Most of the experts (with the important exception of E. Wigner, the founder of the quantum reflection symmetry
notion\footnote{The possible violation of discrete symmetries was actually admitted by Wigner before,
cf. footnote 9 in the famous paper \cite{WWW52} where the notion of superselection sectors was introduced.})
%%(although Wigner didn't remember it in 1982, see \cite{Y82}).
met the possibility of parity violation with reservation,
see e.g. the recollections in the Discussion chapter at the end of \cite{Y82}.
%%C.N. Yang, {\em The discrete symmetries P, T and C}, J. de Physique Colloques {\bf 43 (C8)} (1982) 439-451.
Feynman himself thought of it as possible but unlikely and even proposed later a 50:1 bet against it.

T.D. Lee and C.N. Yang however decided to reconsider the problem performing a thorough examination of the experimental evidence.
Their analysis in \cite{LY} (the paper was received by the editors on June 22 and appeared on October 1, 1956)
showed that $P$-invariance of strong and electromagnetic interactions was confirmed to a high degree of precision
but experiments involving weak interactions "had actually no bearing on the question of parity conservation".
The argument of Lee and Yang was simple: to verify $P$, one needed a pseudoscalar (like the projection
of spin along a momentum, or the mixed product of three momenta) formed out of the measured quantities, and no such
information could be extracted from the data available to that moment (see however Remark 2 below). To fill this gap, Lee
and Yang proposed feasible experimental tests of parity conservation in $\b\,$ decays and, separately, in meson and hyperon decays.

In particular, they argued that a simple possibility to detect $P$-violation would be the asymmetry $\a \ne 0\,$ in the angular distribution
\be
I(\theta)\, d \theta \sim (1 + \a\, \cos \theta ) \sin \theta\, d \theta
\lb{Ib}
\ee
of the $\b\,$ radiation emitted by a {\em polarized} nucleus, say Co$^{60}$.
(Here $\theta\,$ is the angle between the orientation of the nucleus and the electron momentum.)
Obviously, finding that $I(\theta) \ne I(\pi - \theta)\,$ (i.e., $\a \ne 0$) would constitute "an unequivocal proof that
parity is not conserved in $\b\,$ decay".

Noting that without the parity invariance constraint the most general form
of the $\b\,$ decay interaction would involve 10 arbitrary complex constants,
\ba
&&\sum_{i=S,V,T,A,P} (\widetilde p (x)\, O_i\, n (x) )\,(\widetilde e (x)\,
O^i\, (C_i + C'_i \g_5)\, \nu (x)) + h.c.\ ,\nn\\
&&O_S = \id\ ,\ O_V = \g_\mu\ ,\ O_T = \s_{\mu \nu}\ ,\ O_A = \g_\mu \g_5\ ,\ O_P = \g_5\ ,
\lb{LY2}
\ea
Lee and Yang expressed the anisotropy parameter $\a\,$ (\ref{Ib}) in terms of these.

T.D. Lee discussed the cobalt experiment\footnote{The $\b$ decay of
Co$^{60}$ (into an excited state of Ni$^{60}$) is of Gamow-Teller type, i.e. with a spin difference of 1 \cite{Wu57}.}
with Mme C.S. Wu from Columbia University,
%(ground state of 27Co60 is 5+, decaying into the first 4+ state in 28Ni60
a renowned expert in $\b$ decay, even before their paper with Yang was submitted to Physical Review in June 1956 \cite{M}.
Overcoming a series of problems with cryogenics, Wu and her %specially chosen
team from the National Bureau of Standards at Washington obtained
the first result confirming the anisotropy (with $\a \sim - 0.4$, i.e. more electrons in the direction opposite
to the spin of the nucleus) on December 27, 1956.\footnote{A detailed description of the
experiment is contained e.g. in the beautiful exposition by P. Forman \cite{Fo}.}
When the news pointing at parity violation reached the next day Columbia, Leon Lederman who worked with the cyclotron there
decided to perform, together his graduate students M. Weinrich and R. Garwin, an independent test involving
pion and muon decays (another one of those proposed by Lee and Yang). In the first half of January 1957 the preliminary results
of Wu were confirmed, and those of Lederman also showed a distinct parity violation ($\a \sim -\,\frac{1}{3}$)\footnote{Analogous
to Lederman's results were announced also by J.I. Friedman and V.L. Telegdi \cite{FT57}. Another experiment, similar to Wu's but
with {\em positrons} obtained from the $\b^+$ decay of polarized Co$^{58}$, %(into Fe$^{58}$,
carried out in the Kamerlingh Onnes Laboratory in Leiden, Netherlands showed that "positon emission occurs preferably
in the direction of the nuclear spin" \cite{P57}.}.
The articles \cite{Wu57} and \cite{GLW57} (both received on January 15, 1957) appeared in Physical Review on February 15.

The clear experimental evidence of parity non-conservation in weak decays shocked the physical community.
T.D. Lee and C.N. Yang won the 1957 Nobel prize; this had sad effect on their friendship (see \cite{G99}).
Mme Wu was honored much later: in 1978 she became the first winner of the Wolf prize in Physics.
L. Lederman shared the Nobel Prize in 1988 with M. Schwartz and J. Steinberger "for the neutrino
beam method and the demonstration of the doublet structure of the leptons through the discovery of the muon neutrino".

\smallskip

\noindent
{\bf Remark 2. The Cox (1928) and Chase (1930) experiments~} As it turned out, Lee and Yang's claim that no experiment prior to 1956 had provided a test of parity conservation in weak interactions was not correct. In 1928, a team led by R.T. Cox from the New York University
carried out an investigation of double scattering of beta rays \cite{Cox28},
and Cox'es student C.T. Chase continued the experiments with much improved and more definitive techniques \cite{Chase30}.
Their results showing clear evidence for the negative helicity of the beta rays
(implying parity violation in weak interactions) are commented in detail by Lee Grodzins \cite{Gr59}.
The very interesting comments on this subject of E. Wigner in \cite{Y82} also confirm their priority; see as well \cite{Fr79}.
%A. Franklin, The discovery and nondiscovery of parity nonconservation, {\it %Stud. Hist. Phil. Sci.} {\bf 10}:3 (1979) 201-257.}
Most probably, Lee and Yang have overlooked these (too) early papers due to the practical impossibility for their authors
to use the appropriate "keywords": e.g. the purpose announced in R.T. Cox et al. \cite{Cox28} was that
"... it might be of interest to carry out with a beam of electrons experiments analogous to optical experiments in polarization.".

\section{The two-component neutrino}

Alongside with experimentalists, Lee and Yang's paper \cite{LY} of course inspired theorists to look for models
that could incorporate the breaking of parity. Abdus Salam was the first to notice that parity violation may
be related to the vanishing of the neutrino mass (a common belief at that time).
His argument started as follows: the Lagrangian of the free massless neutrino field $\psi^{(\nu)}\,$
($\equiv \nu (x)\,$ (\ref{LY2})), given by (\ref{LDir}) for $m^{(\nu)}=0\,,$ is {\em chiral invariant},
i.e. invariant for the substitution
\be
\psi^{(\nu)} \ \to \ \g_5 \psi^{(\nu)}\quad\Rightarrow\quad \widetilde \psi^{(\nu)} \ \to \ - \widetilde\psi^{(\nu)} \g_5 \ ,\quad
\widetilde \psi^{(\nu)} \psi^{(\nu)} \ \to \ - \widetilde \psi^{(\nu)} \psi^{(\nu)}\ .
\lb{chir-nu}
\ee
One way to secure that no mass term would be produced in neutrino interactions is to require invariance of
the total Lagrangian for (\ref{chir-nu}) while {\it the other fields remain unchanged} \cite{Salam57}.
Similar arguments, exploited further by L.D. Landau \cite{L57} (who predicted exact $CP$-invariance)
and by T.D. Lee and C.N. Yang \cite{LY3} (who already had information about the progress of C.S. Wu's experiment),
revived H. Weyl's $2$-component theory of 1929 \cite{W29} which we will briefly recall.

The Lagrangian (\ref{LDir}) of a free massless Dirac field splits into chiral parts,
\be
{\cal L}^{m=0}_c = - \,\psi^* \b\, \slashed{\partial}\, \psi =
i\, [\, \psi_L^* (\partial_0 + {\boldsymbol\s} {\boldsymbol{\partial}}) \psi_L +
\psi_R^* (\partial_0 - {\boldsymbol\s} {\boldsymbol{\partial}}) \psi_R \, ]\ ,
%%% \quad\psi_{L\atop{R}} = \frac{1\pm\gamma_5}{2} \psi
%%% \nn\\ &&\psi_{L\atop{R}} = \frac{1+\gamma_5}{2} \psi\ , \quad \psi_R=\frac{1-\gamma_5}{2} \psi\ ,
\lb{LDirm0}
\ee
(cf. (\ref{chbas}), (\ref{LR})) implying the Weyl equations for the two-component fields:
\be
(\partial_0 + {\boldsymbol\s} {\boldsymbol{\partial}}) \psi_L = 0 =
(\partial_0 - {\boldsymbol{\s}} {\boldsymbol{\partial}}) \psi_R\ .
\lb{Weyl-eq}
\ee
In contrast with the massive case (\ref{zeta-spin}), the positive and negative frequency
Fourier modes (for each chirality) obey {\em identical} homogeneous equations,
\be
(|{\boldsymbol p}| + {\boldsymbol \s} {\boldsymbol p} ) \, u^\pm_L (p) = 0 \ ,\quad
(|{\boldsymbol p}| - {\boldsymbol \s} {\boldsymbol p} ) \, u^\pm_R (p) = 0\ .
\lb{spinm0}
\ee
The matrices $|{\boldsymbol p}| \pm {\boldsymbol \s} {\boldsymbol p}\,$ being degenerate, Eqs.(\ref{spinm0})
only have {\it one dimensional spaces} of solutions, implying negative/positive {\em helicity}\footnote{Helicity
is the appropriate invariant counterpart of spin in the massless case, see e.g. \cite{BLOT}. By analogy with the screw
rule for circularly polarized electromagnetic waves, negative or positive helicity is equivalent to
"left or right handedness", respectively.} for the corresponding left/right chiral one-particle states of definite momentum,
respectively:
\be
\frac{{\boldsymbol J}.{\boldsymbol p}}{|{\boldsymbol p}|} \, | p\rangle_L =
-\,\frac{1}{2} \, | p\rangle_L\ ,\quad
\frac{{\boldsymbol J}.{\boldsymbol p}}{|{\boldsymbol p}|} \, | p\rangle_R =
\frac{1}{2} \, | p\rangle_R\ ,\qquad
{\boldsymbol J} = \frac{1}{2}\,{\boldsymbol \s}\ .
\lb{chirality}
\ee
In a sense, chiral invariance implies maximal parity violation, as the space inversion transformation of Dirac fields
\be
U(I_s)\, \psi (x) \,U(I_s)^* = \g^0\, \psi (I_s x)\ ,\quad
U(I_s)\, {\widetilde\psi} (x)\, U(I_s)^* = {\widetilde\psi} (I_s x)\, \g_0
\lb{IsD}
\ee
($\g_0 \g^0 = \id\,$) exchanges left and right handed Weyl components:
\be
U(I_s)\, \psi_L (x) \,U(I_s)^* = - i\, \psi_R (I_s x)\ ,\quad
U(I_s)\, \psi_R (x) \,U(I_s)^* = - i\, \psi_L (I_s x)\ .
\lb{IsW}
\ee
The same is valid for the charge conjugation (cf. (\ref{LR})),
\be
U(C)\, \psi_L (x) \,U(C)^* = \psi^*_R (x)\, c^{-1}\ ,\quad
U(C)\, \psi_R (x) \,U(C)^* = \psi^*_L (x)\, c\ ,\quad c = i \s_2
\lb{IsW1}
\ee
which implies, in particular, that the two component (left-handed) neutrino and the corresponding (right-handed)
antineutrino are different particles.

The assumption of chiral invariance (\ref{chir-nu}) (with respect to transformations of the massless neutrino field only)
has important consequences. In their "second" paper Lee and Yang pointed out that it reduces
the number of arbitrary constants in the general parity non-conserving Hamiltonian (\ref{LY2}) to five, since
\be
\widetilde e (x)\, O^i\, (C_i + C'_i \g_5) \frac{1}{2}\,(1+\g_5)\, \nu (x) =
(C_i + C'_i)\, \widetilde e (x)\, O^i\,\nu_L (x)\ .
\lb{LY21}
\ee
They also proposed several new tests of the two component neutrino theory as
e.g. the measurement of {\em momentum and polarization of electrons} emitted in the beta decay of an
{\em unoriented nuclei} which provides another observable pseudoscalar.
Such experiments were promptly executed\footnote{See for example \cite{F57}.} and the results agreed with these of Wu et al.

The overall success of the two component neutrino theory, based on the assumption of an "accidental" (not following from any
gauge principle) vanishing of the neutrino mass, was limited. It could not explain parity violating effects in weak
interactions not involving neutrino which contradicted the idea of universality -- and in particular, it did not help to
resolve the $\theta - \tau$ puzzle.

\smallskip

\noindent
{\bf Remark 3. The reaction of Pauli~}

In 1933 Wolfgang Pauli, the inventor of the neutrino,
had made the following comment on the two component Weyl equations
(\ref{Weyl-eq}): "{\em Indessen sind diese Wellengleichungen, wie ja aus ihrer Herleitung hervorgeht,
nicht invariant gegen\"uber Spiegelungen (Vertauschung von links und rechts) und infolgedessen
sind sie auf die physikalische Wirklichkeit nicht anwendbar.}" [However, as the derivation shows, these wave equations
are not invariant under reflections (interchanging left and right) and thus are not applicable to physical reality.]\footnote{
Page 226 in W. Pauli, Die allgemeinen Prinzipien der Wellenmechanik \cite{P33}; translated as in \cite{Fr79}, p. 215.} An excerpt
from \cite{Fr79} reads: "Even as late as January 17, 1957, Pauli still had not given up his belief. In a
letter to Victor Weisskopf he wrote, '{\em I do not believe that the Lord is a weak left-hander, and I am ready to
bet a very large sum that the experiments will give symmetric results.}' In a letter to C. S. Wu on January 19, 1957,
after hearing word of the results of her experiment he noted, '{\em I did not believe in it (parity nonconservation)
when I read the paper of Lee and Yang.}' " And further: "In another letter to Weisskopf he wrote: '{\em Now, after the first
shock is over, I begin to collect myself. Yes, it was very dramatic. On Monday, the twenty-first, at 8 p.m. I was to give
a lecture on the neutrino theory. At 5 p.m. I received three experimental papers (those of Wu, Lederman, and Telegdi)...
I am shocked not so much by the fact that the Lord prefers the left hand as by the fact that He still appears to be
left-right symmetric when He expresses Himself strongly.}' "

A few months before that (and two and a half years before Pauli's death on 15 December 1958)
the great man's prediction from 1930 of the mere existence of the (anti)neutrino had been finally confirmed by
Frederick Reines and Clyde L. Cowan, Jr. \cite{CR56}. In their experiment based on inverse beta decay,
\be
{\bar\nu}_e + p \to e^+ + n
\lb{nu+p}
\ee
carried near the 700 MW Savannah River nuclear plant, Cowan and Reines detected both the positron and the neutron which
produced $\gamma$-quanta of specific energies after annihilation and delayed capture in cadmium, respectively.
On 14 June 1956 they sent Pauli the following telegram from Los Alamos: "We are happy to inform you that we have definitely
detected neutrinos from fission fragments by observing inverse beta decay of protons.
Observed cross-section agrees well with expected six times ten to minus fourty-four square centimeters.\footnote{The order
$10^{-44}$ cm$^{2}$ of the cross-section was estimated correctly by H. Bethe and R. Peierls already in 1934, in the first
paper on the subject \cite{BP34}. The result was commented pessimistically by the authors who concluded that
"there is no practically possible way of observing the neutrino".} Frederick Reines, Clyde Cowan."

"Thanks for message. Everything comes to him who knows how to wait.", answered on the next day Pauli \cite{Enz}.

\smallskip

A year and a half later M. Goldhaber, L. Grodzins and A.W. Sunyar \cite{GGS}confirmed the left-handedness of the (electron)
neutrino in an ingenious experiment using electron capture by the nucleus Eu$^{152}$:
\be
Eu^{152\, m} + e^- \to Sm^{152\, *} + \nu_e \to Sm^{152} + \g + \nu_e\ .
\lb{Eu}
\ee

\section{The V-A hypothesis}

By the time of the Seventh Rochester Conference (April 15-19, 1957) the fast growing amount of experimental
data from weak decays (including the $\b$-decay and also the decays of $\mu\,, \pi\,$ and strange particles)
started to raise strong concern about the validity of the Universal Fermi Interaction (UFI) hypothesis.
Various decays seemed to favor different combinations of the Lorentz invariants in (\ref{LY2}) so that,
to save UFI, one had to refute some experimental evidence. It was widely believed, in particular,
that the $\b$-decay coupling was S (for Fermi type transitions in nuclei) and T (for Gamow-Teller, i.e. spin changing, transitions)
and the $\mu$-decay coupling, V and A.

The "V-A hypothesis", i.e. the assumption that in fact {\em only the left components $\psi_L\,$ of all four fermion fields}
(and not only those of the neutrino -- and implicitly, of the electron, what concerns the $\b$-decay)
take part in the weak interaction, was made by E.C.G. Sudarshan and R.E. Marshak \cite{SM1, SM2} and independently,
by R. Feynman and M. Gell-Mann \cite{FG-M}. As the Dirac conjugate of $\Pi_L \psi\,$ is
$\widetilde{\psi}\,\Pi_R\,$ where $\Pi_{L\atop{R}} = \frac{1}{2}\,(1\pm\g_5)\,,$ and due to the equalities
\ba
&&[\g_5 , O_i ] = 0 \quad\Rightarrow\quad \Pi_R\, O_i\, \Pi_L = O_i \,\Pi_R \,\Pi_L = 0 \qquad {\rm for}\quad i = S, T, P\ ,\nn\\
%\lb{Oc}
&&[\g_5 , \g_\mu ]_+ = 0 \quad\Rightarrow\quad \Pi_R\, O_i\, \Pi_L = O_i \,\Pi_L^2 = O_i \,\Pi_L \qquad {\rm for}\quad i = V, A\ ,\nn\\
%\lb{Occ}
&&\g_\mu\, \Pi_L\otimes \g^\mu \, (C_V + C'_V \g_5)\, \Pi_L +
\g_\mu \g_5 \, \Pi_L\otimes \g^\mu \g_5 \,(C_A + C'_A \g_5)\, \Pi_L =\nn\\
&&= (C_V+C'_V+C_A+C'_A)\, \g_\mu \, \Pi_L \otimes \g^\mu \, \Pi_L\ ,
\lb{Oc1}
\ea
see (\ref{LY2}), the V-A hypothesis amounts to setting
\be
C_i = C'_i = 0\ ,\quad i = S, T, P,\qquad C_V = C'_V = C_A = C'_A \equiv \frac{G}{4\sqrt{2}}\ .
\lb{Oc2}
\ee
The number of the coupling constants is thus reduced to one and the UFI Hamiltonian takes the following simple and elegant form:
\be
\frac{G}{\sqrt{2}}\, (\widetilde \psi_4 (x)\, \g_\mu \Pi_L\, \psi_3 (x) )\,(\widetilde \psi_2 (x)\,
\g^\mu \Pi_L\, \psi_1 (x)) + h.c.\ .
\lb{LY3}
\ee

V-A was in a very good agreement with most of the experiments but its authors had to make the bold assumption that
the predictions of the rest were actually erroneous; needless to say, this required a thorough work.
Sudarshan and Marshak \cite{SM1} concluded: "{\it While it is clear that a mixture of vector and axial
vector is the only universal four-fermion interaction which is possible and possesses many elegant features, it appears
that one published \cite{RR53}\footnote{Sudarshan and Marshak cited also the later paper \cite{RR55}.}
and several unpublished experiments cannot be reconciled with this hypothesis...
All of these experiments should be redone... If any of the above four experiments stands, it will be necessary to
abandon the hypothesis of a universal V+A four-fermion interaction...}"\footnote{Sheldon Glashow commented on this in
\cite{G09}: "{\it This is theoretical physics at its zenith! The experiments were redone with results that now
confirmed their hypothesis. It was a stunning accomplishment, yet one which has never been recognized with a prize.}"}.

In their review \cite{G-MR} published in December 1957 (the survey of literature pertaining to the
review being completed in July, 1957) Gell-Mann and Rosenfeld made the following comment:
"{\it ... there has been speculation that the form of the interaction might also be 'universal'.
Such a situation seems to be ruled out if the $\b$-decay coupling is primarily S and T
and the $\mu\,$ decay coupling V and A... Since the $\b$-decay picture is somewhat confused at the moment,
let us discuss briefly the possibility that we may have V and A there too, instead of S and T with a possible admixture of V.
We may call this V, A hypothesis the 'last stand' of the UFI... We must first of all disregard much of the evidence on
$e - \nu\,$ angular correlation in $\b$-decay, especially the result of Rustad} \& {\it Ruby\footnote{In \cite{RR55}.}
on He$^6$, which clearly indicates T rather than A. This is already a very serious objection to the UFI.}"

Discussing the universality of the weak interaction with V-A in \cite{FG-M}, Feynman and Gell-Mann also stressed that
"{\it ... At the present time several $\b$-decay experiments seem to be in disagreement with one another.
Limiting ourselves to those that are well established, we find that the most serious disagreement with our
theory is the recoil experiment in He$^6$ of Rustad and Ruby\footnote{Again, \cite{RR55}.}
indicating that the T interaction is more likely than the A. Further check on this is obviously very desirable."}
Feynman's picturesque story in \cite{FL, Me} about experimental points at the edge of the data range reflects his
personal battle with this problem.

\smallskip

%Leaving aside for a while the question of priority, perhaps the most intriguing part of the V-A saga was the fact
%that the identical conclusions reached in \cite{FG-M} and \cite{SM1, SM2} originated from quite different sources of inspiration.

Feynman's line of thought actually started with the following intriguing observation, documented first in the Proceedings of the
Seventh Rochester Conference \cite{F-VII}, and then in \cite{FG-M}. The electron wave function in the presence of
{\em electromagnetic} interaction\footnote{See the Appendix for the free case conventions adopted here.}
obeys the Dirac equation
\be
(\sD + m) \psi = 0  \qquad (\, \sD = \g^\mu D_\mu\ ,\quad D_\mu = \partial_\mu + i e\, A_\mu\ ,\quad F_{\mu\nu} =
\partial_\mu A_\nu - \partial_\nu A_\mu )\ .
\lb{FDir1}
\ee
Multiplying it by $\sD - m\,,$ one obtains
\be
\sD^2 \psi = m^2\, \psi \qquad (\,\sD^2 = D^\mu D_\mu  + e\, \frac{i}{2}\, \s^{\mu\nu} F_{\mu\nu}\ ,\quad\s^{\mu\nu} =
\frac{i}{2}\, [\g^\mu , \g^\nu] \, )\ .
\lb{FDir2}
\ee
The key point here is that $[\g_5 , \s^{\mu\nu} ] = 0\,$ so that any solution of (\ref{FDir2}) can be presented as a
sum $\psi = \psi_+ + \psi_-\,$ where $\psi_\pm\,$ also solve (\ref{FDir2}) and
\be
\g_5 \, \psi_\pm = \pm \psi_\pm\qquad \Leftrightarrow\qquad
\frac{1}{2}\,(1\pm \g_5) \, \psi_\pm = \psi_\pm = \frac{1}{2}\,(1\pm \g_5) \, \psi\ .
\lb{FDir21}
\ee
Clearly, both $\psi_+\,$ and $\psi_-\,$ only have two independent components.
It is most appropriate to use the chiral basis (\ref{chbas}) in which $\psi_+ = \begin{pmatrix} \psi_L\cr 0\end{pmatrix}\,,\
\psi_- = \begin{pmatrix} 0 \cr\psi_R \end{pmatrix}\,$ and the matrix part of (\ref{FDir2}) is block diagonal:
\ba
&&\frac{i}{2}\, \s^{\mu\nu} F_{\mu\nu}
= F_{0k} \begin{pmatrix} - \s_k&0\cr 0&\s_k\end{pmatrix}
- \frac{i}{2}\, \varepsilon_{ijk} F_{ij} \begin{pmatrix} \s_k&0\cr 0&\s_k\end{pmatrix} = \lb{FDir3}\\
&&= - \begin{pmatrix} (\boldsymbol{E} + i \boldsymbol{B}) .
{\boldsymbol{\s}}&0\cr 0& (- \boldsymbol{E} + i \boldsymbol{B}) .
{\boldsymbol{\s}} \end{pmatrix} \ ,\quad E_k = F_{0k}\ ,\quad B_k
= \frac{1}{2}\, \varepsilon_{ijk} F_{ij}\ . \nn \ea As a result,
the two-component spinors $\psi_L\,$ and $\psi_R\,$ satisfy
separate Klein-Gordon type equations only involving Pauli
matrices; choosing the first one we obtain \be [ (\partial^\mu + i
e\, A^\mu) ( \partial_\mu + i e\, A_\mu ) - e\,(\boldsymbol{E} + i
\boldsymbol{B}) . {\boldsymbol{\s}}]\, \psi_L = m^2\, \psi_L\ .
\lb{FKG} \ee Feynman "has always had a predilection" for
Eq.(\ref{FKG}) because "{\it if one tries to represent
relativistic quantum mechanics by the method of path integrals,
the Klein-Gordon equation is easily handled, but the Dirac
equation is very hard to represent directly}" \cite{FG-M}. He had
checked that the rules of calculation for electrodynamics worked
out on the basis of (\ref{FKG}) give exactly the same results as
those calculated with Dirac matrices. Moreover, he argued, to
describe the spin 1/2
%$\frac{1}{2}\,$
of the electron and the positron one only needs two component
(Pauli) spinors, if one uses a second order equation. (Similarly,
for a spin $0\,$ particle the single component Klein-Gordon
equation automatically takes into account the positive and
negative energy solutions, the latter being specified by initial
conditions for both the function and its time derivative.) Now
having secured the point that the chiral projection makes sense
even in the massive case, Feynman suggested %the rule
that in the $\b$-decay Hamiltonian the electron (instead of the
massless neutrino!) field $\psi^{(e)}\,$ should be replaced by
$\frac{1}{2}\,(1+\g_5) \, \psi^{(e)}\,.$ So the final steps in
\cite{FG-M} towards V-A -- putting forward the hypothesis that
"{\it the same rule applies to the wave functions of all the
particles entering the interaction}" (cf. (\ref{Oc1})) and then
assuming universality of the coupling, looked already quite
natural for him.

\smallskip

Sudarshan and Marshak \cite{SM1, SM2} derived (\ref{LY3}) instead by directly requiring the invariance of the weak interaction
Hamiltonian with respect to chiral transformations of any of the four fermion fields separately. Clearly, this is a universal
condition, stronger than the invariance (\ref{chir-nu}) which is only applicable when one of the fields is a (massless) neutrino.
The explicit calculation was provided actually soon after in \cite{S57} by J.J. Sakurai who included chiral invariance into a
more general framework, generalizing the earlier observation of J. Tiomno \cite{T55} that
the Dirac equation (\ref{FDir1}) is invariant under simultaneous chiral {\em and mass reversal} transformations:
\be
\psi \to \eta\, \g_5 \psi\ ,\quad \widetilde \psi \to - \eta^* \widetilde \psi \,\g_5 \quad
( |\eta|^2 = 1)\ ,\quad m \to - m\ , \quad A_\mu \to A_\mu \ .
\lb{Tiomno}
\ee
(It is worth mentioning that, in the free case, inverting the sign of the mass is equivalent to
exchanging $u_\zeta (p) \leftrightarrow v_\zeta (p)\,,$ see (\ref{DirSln}) and (\ref{zeta-spin}). %and (\ref{sln-norm}).
Clearly, (\ref{Tiomno}) is also a symmetry of Feynman's second order formulation of QED based on (\ref{FDir2}) and (\ref{FKG}) since it
only contains the square of $m\,.$) In effect, both the electromagnetic and the weak Lagrangians (including the mass
terms) are left invariant with respect to the combined chiral transformation and mass inversion.

\medskip

In 1957, both R.P. Feynman (May 11, 1918 -- February 15, 1988) and
the younger M. Gell-Mann (born September 15, 1929) were already
well established at Caltech, while E.C.G.
Sudarshan\footnote{Ennackal Chandy George Sudarshan (16 September
1931 -- 14 May 2018) was born in Pallom, a small village in the
Kerala state, south India.} was finishing his PhD thesis under the
supervision of R.E. Marshak (October 11, 1916 -- December 23,
1992), then chairman of the Physics Department at Rochester. In
Marshak's recollections from 1991 \cite{M91}, "{\it It was
therefore completely natural -- after the Wu et al. announcement
-- to suggest to Sudarshan that he might take a fresh look at to
whether a common Lorentz structure and strength could be assigned
to all weak interactions. ... Sudarshan plunged into this problem
with alacrity and exceeding good taste... By the time of the
Seventh Rochester Conference in April 1957, it was clear to both
Sudarshan and myself that the only possible UFI for weak processes
was V-A (with a lefthanded neutrino) and not a combination of S
and T (with a righthanded neutrino), as was widely believed.}".
The first official announcement on the way to V-A was made however
by Feynman (who borrowed a few minutes from Kenneth Case's talk
time for a quick outline of his ideas at Rochester VII \cite{Me})
while Marshak did not opt to talk on theirs. Informally, this has
been done at a luncheon meeting in the first week of July at
Caltech (attended by Sudarshan, Marshak, Gell-Mann as well by the
experimentalists F. Boehm and A.H. Wapstra, the theoretician B.
Stech and some others from Caltech); a few days later the abstract
of \cite{SM1} has been sent to Prof. N. Dallaporta, chairman of
the Padua-Venice Conference on Mesons and Newly Discovered
Particles, 22-27 September 1957. The preprints of \cite{SM1} and
\cite{FG-M} both appeared on September 16, 1957 \footnote{Exactly
on the 26-th birthday of E.C.G. Sudarshan.}. However, while
\cite{SM1}, after having been reported at the meeting in Italy,
had to await publication till May 2018 (only to be practically
buried in the proceedings), Feynman and Gell-Mann's paper was sent
immediately to Physical Review where it appeared on the first day
of 1958. On January 10, 1958 Sudarshan and Marshak sent a short
letter to the same journal which appeared in the March 1 issue.
Their intention was "{\it to take stock of experimental
developments favoring the V-A theory since the Padua-Venice
Conference}" \cite{M91}; in Marshak's own words, "{\it Our short
note ... was not intended as a substitute for our detailed 1957
Padua-Venice paper but, unfortunately, it was treated by all too
many physicists in later years as the sole publication of our V-A
theory.}".

The personal viewpoint of E.C.G. Sudarshan and R.E. Marshak on the
emergence and the public reception of the V-A theory is presented
e.g. in \cite{SM3, SM4} as well in the aforementioned Marshak's
1991 talk "The pain and joy of a major scientific discovery"
\cite{M91} on the occasion of E.C.G. Sudarshan's 60th
birthday\footnote{In \cite{M91}, one year before his death in an
accidental drowning in Cancun, Mexico, Marshak admitted "three
cardinal blunders" of his that prevented Sudarshan to receive the
due share of fame for the V-A discovery.}. R.P. Feynman's feelings
about his own participation in these events are reflected in "The
7 percent solution" article in \cite{FL} and, even stronger, in
the interview given to Jagdish Mehra in January 1988, shortly
before his death \cite{Me}: "{\it As I thought about it, as I
beheld it in my mind's eye, the goddamn thing was sparkling, it
was shining brightly! As I looked at it, I felt that it was the
first time, and the only time in my scientific career, that I knew
a law of nature that no one else know. Now, it wasn't as beautiful
a law as Dirac's or Maxwell's but my new equation for beta decay
was a bit like that. It was the first time that I discovered a new
law, rather than finding a more efficient method of calculating
from someone else's theory ... I learned later that others had
thought of it at about the same time or a little before, but that
didn't make any difference. At the time I was doing it, I felt all
the thrill of a new discovery! ... I thought, 'Now I have
completed myself!'}". Yet, after years have passed, according to
Mehra Feynman was "perfectly happy to share the credit for the
discovery with Gell-Mann, Sudarshan, and Marshak.". The role of M.
Gell-Mann in V-A is justly described by G. Johnson in Chapter 7 "A
Lopsided Universe" of the biographical book \cite{J00}.

\smallskip

The incontestable experimental confirmation, by the end of 1957,
of the neutrino left-handedness \cite{GGS} and the careful
re-examination and correction, in the following couple of years,
of all the data pointed in \cite{SM1, SM2} and \cite{FG-M} as
contradicting the V-A theory, gave the latter a triumphant status.
It is quite impressive to read the retrospective judgment of two
of the greatest living particle theorists, the Nobel laureates S.
Glashow \cite{G09} and S. Weinberg \cite{W09} presented more than
a half a century later at the Sudarshan Symposium in 2009. For
one, it is for the priority issues commented by them from the
standpoint of direct participants, both as young scientists just
beginning their carrier in most glorious days of V-A and then, as
true leaders in the following developments. Not less important is
the value assigned by both of them to the achievement; in
Weinberg's words, "V-A was the key" to the future.

\section{The end of the beginning}

In a sense, establishing the V-A structure of the weak Hamiltonian
marked the end of the beginning. The following is well known: in a
nutshell, it included the idea of electroweak unification as a
$U(1) \times SU(2)\,$ gauge theory, the Brout-Englert-Higgs
%(-Guralnik-Hagen-Kibble)
mechanism and the proof of the renormalizability of the ensuing
theory; the latter was backed by the discovery of its main
ingredients, the neutral currents, the massive intermediate bosons
and finally, the Higgs boson, in a series of spectacular
experiments, most of which have been carried at CERN. Several
Nobel prizes have been awarded in recognition of the progress,
both theoretical and experimental.

Some of the properties of neutrinos remain still elusive, and the quest for clarification is at the forefront of contemporary particle physics. B. Pontecorvo's idea from 1957-1958 of neutrino oscillations turned out to be correct, showing in particular that neutrinos, after all, are massive.

The parity violation and the resulting V-A theory taught us that
the basic players in QFT are irreducible representations of the
simply connected quantum mechanical Lorentz group $SL(2,
\mathbb{C})$. The only surviving discrete symmetry, Pauli's
$CPT\,$ theorem, is a consequence of continuous space-time
symmetries combined with the basic principles of local QFT.

Weak interactions are unique among all known (four, including
gravity) in two respects -- they violate parity and are mediated
by massive bosons. From its emergence to its ultimate success,
their story added lots of examples verifying T.D. Lee's two laws
of physicists: "Without experimentalists, theorists tend to drift.
Without  theorists, experimentalists tend to
falter."\footnote{T.D. Lee, History of the weak interactions, Talk
at the "Jackfest" marking the 65th birthday of Jack Steinberger,
see e.g. CERN Courier, January/February 1987.}

\section*{Acknowledgments}
\addcontentsline{toc}{section}{Acknowledgments}

The author thanks Ivan Todorov and Serguey Petcov for their precious suggestions and critical reading of the manuscript.
Without their help and encouragement this survey simply wouldn't exist. This work was supported in part by the Bulgarian
National Science Fund under research grant DN-18/3.

\newpage

\section*{Appendix.\\ Gamma matrices and the free spin $1/2\,$ field}
\addcontentsline{toc}{section}{Appendix. Gamma matrices and the free spin $1/2\,$ field}

\setcounter{equation}{0}
\renewcommand\theequation{A.\arabic{equation}}

We will use the {\it space-like} metric $\eta_{\mu\nu}\,$ in Minkowski space so that e.g.
\be
p^2 = p^\mu \eta_{\mu\nu} p^\nu = p_\mu\, \eta^{\mu\nu} p_\nu =
{\mathbf p}^2 - p_0^2\ ,
\qquad \bp^2 = p_1^2 + p_2^2 + p_3^2\ .
\lb{Mm}
\ee
The Clifford algebra $C\ell\, (3,1)\,$ generated by the $\g\,$ matrices satisfying
\be
[\gamma_\mu , \gamma_\nu ]_+ := \g_\mu \g_\nu + \g_\nu \g_\mu = 2\,\eta_{\mu \nu}\ ,\qquad (\eta_{\mu\nu}) = \rm{diag} (-,+,+,+)
\lb{conv-gamma}
\ee
is isomorphic to the algebra of $4\times 4\,$ {\it real}
matrices\footnote{A real (Majorana) representation is given e.g.
by
$\g_0^M = \s_3\otimes c\ ,\quad \g_1^M = \id\otimes \s_3\ , \quad\g_2^M = \id\otimes \s_1\ ,\quad\g_3^M = - c \otimes c\quad \ ( c = i\,\s_2 )\ ;\quad\
\g_0^M \g_1^M \g_2^M \g_3^M = \s_1 \otimes c = - (\s_1 \otimes c)^t\ .$}, see e.g. \cite{T11}.
The Dirac conjugate of a $4$-component spinor $\psi\,$ is defined as
\be
{\widetilde \psi} = \psi^* \b\ ,\qquad\g^*_\mu\, \b = - \b\, \g_\mu %%\qquad \beta=i\g^0=\beta^*\ .
\lb{Dir-conj}
\ee
(the star $^*$ standing for hermitean conjugation) and the charge conjugate, as
\be
\psi^C = {\widetilde\psi}\, C^{-1} \ ,\qquad \gamma^t_\mu C = -\, C \,\gamma_\mu %%\equiv  \psi^* B \ ,\qquad B = \b\,C^{-1}\ .
\lb{Cspin}
\ee
where $C\,$ is the charge conjugation matrix and $^t$ stands for transposition\footnote{The matrix $\b\,$ defines a hermitean form, and $C = (C_{\a\b})\,$ a skew-symmetric bilinear form on the spinors $\psi = (\psi^\a)\,.$}.
It is convenient to define $\g_5\,$ as a hermitean matrix:
\ba
&&\g_5=i\, \g_0\g_1\g_2\g_3 \ ;\qquad {\rm tr}\,\g_5 \g^\mu\g^\nu\g^\rho\g^\sigma =
4i\,\epsilon^{\mu\nu\rho\sigma} \quad (\epsilon^{0123} = 1 = -\epsilon_{0123}) \ .\qquad\qquad
\lb{tr-g}
\ea
The chiral ($\g_5$-diagonal) basis of $\g$ matrices can be realized as
\ba
&&i \g^0 = \begin{pmatrix}0&\id\cr \id&0 \end{pmatrix} = \b\ ,\quad
i \boldsymbol{\g} = \begin{pmatrix}0&\boldsymbol{\s}\cr -\boldsymbol{\s}&0 \end{pmatrix}\ ,\qquad
\g_5 = %i\, \g_0 \g_1 \g_2 \g_3 =
\begin{pmatrix}\id&0\cr 0&- \id \end{pmatrix}\ \nn\\
&&C = \begin{pmatrix} c&0\cr 0&c^{-1} \end{pmatrix} \ ,\qquad c = i\, \s_2 = \begin{pmatrix} 0&1\cr -1&0\end{pmatrix}
= - c^t = - c^{-1}\ .
\lb{chbas}
\ea

\medskip

The Lagrangian of a free (complex) Dirac field is
\ba
&&{\cal L}= \frac{1}{2}\, (\partial_\mu {\widetilde \psi}\, \g^\mu - {\widetilde \psi}\,
\slashed{\partial})\, \psi - m \, {\widetilde \psi}\, \psi =
{\cal L}_c +\frac{1}{2}\,\partial_\mu ({\widetilde \psi}\,\g^\mu\, \psi)\ ,\nn\\
&&{\cal L}_c = - \,{\widetilde \psi}\, (m + \slashed{\partial} )\, \psi\ ,\qquad
\slashed{\partial} = \g^\mu \partial_{\mu} \equiv\g^\mu\frac{\partial}{\partial x^\mu}\ ,
\lb{LDir}
\ea
so that the Dirac equation for $\psi (x)\,$ and the conjugate equation for ${\widetilde\psi} (x)\,$ read
\be
(\slashed{\partial} + m)\, \psi (x) = 0\ ,\qquad \partial_\mu {\widetilde \psi}(x) \g^\mu - m\,{\widetilde \psi}(x) = 0\ .
\lb{DirEoM}
\ee
The solution of the Dirac equation (\ref{DirEoM}) is presented in the form
\ba
&&\psi (x) = \sum_\zeta \int [ b_\zeta (p, +)\, e^{ipx}\, u_\zeta(p) + b^*_\zeta (p, -)\, e^{-ipx}\, v_\zeta(p)]\, (dp)_m\ ,\nn\\
&&{\widetilde\psi}(x) = \sum_\zeta \int [ b^*_\zeta (p, +)\, e^{-ipx}\, {\widetilde{u}}_\zeta(p) +
b_\zeta (p, -)\, e^{ipx}\, {\widetilde{v}}_\zeta(p) ]\, (dp)_m \ ,\quad
\lb{DirSln}
\ea
where the invariant measure on the future mass hyperboloid $(dp)_m\,$ is
\be
(2\pi)^3(dp)_m=[\int_0^\infty \delta(p^2+m^2)\, dp_0]\, d^3p = \frac{d^3p}{2\omega_{\mathbf p}}\ ,\quad
\omega_{\mathbf p} \, ( =\vert p_0 \vert ) = \sqrt{m^2+{\mathbf p}^2}\ .
\label{dpm}
\ee
Here $u_\zeta\,$ and $v_\zeta\,$ are classical ("c-number") spinors satisfying the linear algebraic equations
\ba
&&(m + i\, \slashed{p} ) \, u_\zeta (p) = 0 = {\widetilde u}_\zeta (p)\, (m + i\, \slashed{p})\ ,\nn\\
&&(m - i\, \slashed{p} ) \, v_\zeta (p) = 0 =\, {\widetilde v}_\zeta (p)\, (m - i\, \slashed{p})\qquad {\rm for}
\quad p^0 = \o_{\mathbf p}
\lb{zeta-spin}
\ea
and $\zeta= \pm \frac{1}{2}\,$ is the spin projection index.
In the $\g_5$-diagonal basis the field $\psi (x)\,$ (as well as $u_\zeta(p)\,$ and $v_\zeta(p)$) splits into chiral components:
\ba
\lb{LR}
&&\psi(x) = \left( {\psi_L (x)}\atop{\psi_R (x)} \right)\ ,\qquad
\widetilde{\psi}(x) = (\psi_R^* (x) , \psi_L^* (x))\ , \nn\\
&&\psi^C(x) = \widetilde{\psi}(x) C^{-1} = \left( {\psi^*_R (x)\, c^{-1}}\atop{\psi_L^*(x)\, c}\right)\ .
\ea

A Majorana field is equal to its charge conjugate, $\psi (x) = \psi^C(x)\,$ (which amounts in the chiral basis
to $\psi_R (x) = \psi_L^*(x)\, c$). It is a special case of (\ref{DirSln}) with
$b_\zeta (p, +)= b_\zeta (p, -)\,$ and $v_\zeta (p) = u^C_\zeta(p)\,.$


\newpage

\begin{thebibliography}{000}
\addcontentsline{toc}{section}{References}

\bibitem[A84]{A84}
E. Amaldi,
From the discovery of the neutron to the discovery of nuclear fission,
{\it Phys. Rep.} {\bf 111}:1-4 (1984) 1-332 (see Ref. [277] for the origin of the name "neutrino").
%% http://www.hep.princeton.edu/~mcdonald/examples/EP/amaldi_pr_111_1_84.pdf

\bibitem[BP34]{BP34}
H. Bethe, R. Peierls,
The "Neutrino",
{\it Nature} {\bf 133} (1934) 532.

\bibitem[BLOT]{BLOT}
N.N. Bogolubov, A.A. Logunov, A.I. Oksak, I.T. Todorov,
{\it General Principles of Quantum Field Theory},
Kluwer, Dordrecht et al. 1990.

\bibitem[BLT]{BLT}
N.N. Bogolubov, A.A. Logunov, I.T. Todorov,
\textit{Axiomatic Quantum Field Theory},
authorized translation from the Russian manuscript,
Ed. by Stephen A. Fulling, Addison-­Wesley/W.A. Benjamin, 1975, 708 p.

\bibitem[B30]{B30}
N. Bohr, 
Faraday Lecture delivered on May 8th, 1930:
Chemistry and the quantum theory of atomic constitution, 
{\it J. Chem. Soc.}
%(Resumed)
(1932) 349-384.

\bibitem[Ch32]{Ch32}
J. Chadwick, Possible existence of a neutron,
{\it Nature} {\bf 192} (Feb 27, 1932) 312;

J. Chadwick, The Existence of a Neutron,
{\it Proc. Roy. Soc. London Ser. A} {\bf 136} No. 830 (June 1, 1932) 692-708.

\bibitem[Chase30]{Chase30}
C.T. Chase,
The scattering of fast electrons by metals. II. Polarization by double scattering at right angles,
{\it Phys. Rev.} {\bf 36} (1930) 1060-1065.

\bibitem[Cox28]{Cox28}
R.T. Cox, C.G. McIlwraith, B.Kurrelmeyer,
Apparent evidence of polarization in a beam of $\b$-rays,
{\it Proc. Natl. Acad. Sci. USA} {\bf 14} (1928) 544-549.

\bibitem[CR56]{CR56}
C.L. Cowan, Jr., F. Reines, F.B. Harrison, H.W. Kruse, A.D. McGuire,
Detection of the free neutrino: a confirmation,
{\it Science} {\bf 124} (1956) 103-104.

\bibitem[Enz]{Enz}
Ch.P. Enz,
\textit{No Time to be Brief. A scientific biography of Wolfgang Pauli},
Oxford University Press 2002.

\bibitem[F33]{F33}E. Fermi,
Versuch einer Theorie der $\b$-Strahlen. I,
%Preliminary note,
{\it La Ricerca scientifica} {\bf 2}, Heft 12 (1933);
{\it Z. Phys.} {\bf 88}:3 (1934) 161-177
[for a complete translation in English, see
F.L. Wilson, Fermi's theory of beta decay, {\it Amer. J. Phys.} {\bf 36}:12 (1968) 1150-1160];
An attempt to a $\b\,$ rays theory, {\it Il Nuovo Cimento} Nuova Serie N. {\bf 1} (1934) 1-20.

\bibitem[F-VII]{F-VII}
R. Feynman,
Alternative to the 2-component neutrino theory, {\bf in}:
Proceedings of the Seventh Rochester Conference on High Energy Nuclear Physics, 15-19 April 1957,
Interscience, New York, Session IX, pp. 42-44.

\bibitem[FL]{FL}
R.P. Feynman,
{\it "Surely You're Joking, Mr. Feynman!"}.
Adventures of a Curious Character as told to Ralph Leighton, E. Hutchings (ed.),
W.W. Norton, 1985 (available electronically);
see, in particular, "The 7 percent solution" in Part 5 "The World of One Physicist", pp. 161-166.

\bibitem[FG-M]{FG-M}
R. Feynman, M. Gell-Mann,
Theory of the Fermi interaction,
{\it Phys. Rev.} {\bf 109}:1 (1958) 193-198.

\bibitem[Fo]{Fo}
P. Forman,
The Fall of Parity, \href{https://www.nist.gov/pml/fall-parity}{\footnotesize{\tt https://www.nist.gov/pml/fall-parity}}.

\bibitem[Fr79]{Fr79}
A. Franklin,
The discovery and nondiscovery of parity nonconservation,
{\it Stud. Hist. Phil. Sci.} {\bf 10}:3 (1979) 201-257.

\bibitem[F57]{F57}
H. Frauenfelder et al.,
Parity and the polarization of electrons from Co$^{60}$,
{\it Phys. Rev.} {\bf 106}:2 (1957) 386-387, received on March 1, 1957.

\bibitem[FT57]{FT57}
J.I. Friedman, V.L. Telegdi,
Nuclear emulsion evidence for parity nonconservation in the decay chain $\pi^+ - \mu^+ - e^+$,
{\it Phys. Rev.} {\bf 105}:5 (1957) 1681-1682, received on January 17, 1957.

\bibitem[GT36]{GT36}
G. Gamov, E. Teller,
Selection rules for the $\b$-disintegration,
{\it Phys. Rev.} {\bf 49}:12 (1936) 895-899.

\bibitem[GLW57]{GLW57}
R.L. Garwin, L.M. Lederman, M. Weinrich,
Observations of the failure of conservation of parity and charge conjugation in meson decays: the magnetic moment of the free muon,
\textit{Phys. Rev.} \textbf{105}:4 (1957) 1415-1417.

\bibitem[G-MR]{G-MR}
M. Gell-Mann, A.H. Rosenfeld,
Hyperons and heavy mesons (systematics and decay),
{\it Annual Review of Nuclear Science}, vol. {\bf 7} (1957) 407-478
(Volume publication date December 1957; the survey of literature pertaining to the review completed in July, 1957).

\bibitem[G99]{G99}
J. Glanz,
What fuels progress in Science? Sometimes, a feud,
{\it The New York Times}, Science, September 14, 1999.

\bibitem[G09]{G09}
S. Glashow,
Message for Sudarshan Symposium,
{\it J. Physics: Conf. Ser.} {\bf 196} (2009) 011003.

\bibitem[GGS]{GGS}
M. Goldhaber, L. Grodzins, A.W. Sunyar,
Helicity of neutrinos,
{\it Phys. Rev.} {\bf 109} (1958) 1015-1017, received on December 11, 1957.

\bibitem[Gr59]{Gr59}
L. Grodzins,
The history of double scattering of electrons and evidence for the polarization of beta rays,
{\it Proc. Natl. Acad. Sci. USA.} {\bf 45}:3 (1959) 399-405.

\bibitem[H32]{H32}
W. Heisenberg, 
\"Uber den Bau der Atomkerne. I, {\it Z. Phys.} {\bf 77}:1-2 (1932) 1-11; 
II, {\it Z. Phys.} {\bf 78}:3-4 (1932) 156-164.

\bibitem[H1820]{H1820}
J.F.W. Herschel,
On the rotation impressed by plates of rock crystal on the planes of polarization of the rays of light,
as connected with certain peculiarities in its crystallization,
{\it Transactions of the Cambridge Philosophical Society} {\bf 1} (1820) 43-51.

\bibitem[J00]{J00}
G. Johnson,
{\it Strange Beauty (Murray Gell-Mann and the Revolution in Twentieth-Century Physics)},
Vintage Books, New York 2000.

\bibitem[K2018]{K2018}
KATRIN experiment home page:
\href{https://www.katrin.kit.edu/}{https://www.katrin.kit.edu/}.

\bibitem[L57]{L57}
L.D. Landau,
On the conservation laws for weak interactions,
{\it Nucl. Phys.} {\bf 3} (1957) 127-131,
received on January 9, 1957.

\bibitem[L24]{L24}
O. Laporte,
Die Struktur des Eisenspektrums,
{\it Z. Physik} {\bf 23} (1924) 135-175.

\bibitem[LRY49]{LRY49}
T.D. Lee, M. Rosenbluth, C.N. Yang,
Interaction of mesons with nucleons and light particles,
{\it Phys. Rev.} {\bf 75} (1949) 905.

\bibitem[LY2]{LY2}
T.D. Lee, C.N. Yang,
Mass degeneracy of the heavy mesons,
{\it Phys. Rev.} {\bf 102}:1 (1956), 290-291,
received on 29 December 1955, published on 1 April 1956.

\bibitem[LY]{LY}
T.D. Lee, C.N. Yang,
Question of parity conservation in weak interactions,
{\it Phys. Rev.} {\bf 104}:1 (1956) 254-258,
received on June 22, 1956, published on October 1, 1956.

\bibitem[LY3]{LY3}
T.D. Lee, C.N. Yang,
Parity nonconservation and a two-component theory of the neutrino,
{\it Phys. Rev.} {\bf 105}:5 (1957) 1671-1675,
received on January 10, 1957; revised manuscript received on January 17, 1957.

\bibitem[L09]{L09}
A. Lesov,
The weak force: from Fermi to Feynman,
{\tt arXiv:0911.0058 [physics.hist-ph]}.

\bibitem[Ma]{Ma}
J. Magueijo,
{\it A Brilliant Darkness. The Extraordinary Life and Disappearance of Ettore Majorana, the Troubled Genius of the Nuclear Age},
Perseus, Basic Books, N.Y. 2009.

\bibitem[M91]{M91}
R.E. Marshak,
The pain and joy of a major scientific discovery,
Banquet talk on the occasion of E.C.G. Sudarshan's 60th birthday celebration,
{\em Z. Naturforsch.} {\bf 52a} (1997) 3-8.

\bibitem[Me]{Me}
J. Mehra,
\textit{The Beat of a Different Drum: The Life and Science of Richard Feynman},
Oxford Univ. Press, 1994, 630 p.; see, in particular, Section 21 "
'The only law of nature I could lay a claim to': the theory of weak interactions", pp. 453-481.

\bibitem[M]{M}
Krishna Myneni,
Symmetry destroyed: the failure of parity (1984), %December 10, 1984,
\href{https://www.hep.ucl.ac.uk/~nk/teaching/PH4442/parity-violation.html}{\footnotesize{\tt
https://www.hep.ucl.ac.uk/$\sim$nk/teaching/PH4442/parity-violation.html}}.

\bibitem[P30]{P30}
W. Pauli,
Open letter to the group of radioactive people at the
Gauverein meeting in T\"ubingen (Z\"urich, December 4, 1930);
the original text in German and the translation (by Kurt Riesselmann) are available at
\href{http://microboone-docdb.fnal.gov/cgi-bin/RetrieveFile?docid=953;filename=pauli%20letter1930.pdf}{\footnotesize{\tt
http://microboone-docdb.fnal.gov/cgi-bin/RetrieveFile?docid=953;
filename=pauli\%20letter1930.pdf}}.

\bibitem[P33]{P33}
W. Pauli, Die allgemeinen Prinzipien der Wellenmechanik, {\bf in}:
{\it Quantentheorie, Handbuch der Physik} {\bf 24} (1933) 83-272,
H. Bethe et al. (eds.), Springer Berlin, Heidelberg.

\bibitem[P94]{P94}
S.T. Petcov,
On B. Pontecorvo contributions to weak interaction and neutrino physics,
{\bf in}: Proceedings of the 6th International Symposium on Neutrino Telescopes, Istituto Veneto di Scienze, Lettere ed Arti, Venice,
22-24 February, 1994, M. Baldo-Ceolin (ed.), pp. 17-26; \\ available at 
\href{https://cds.cern.ch/record/265610/files/P00024340.pdf}
{\footnotesize{\tt https://cds.cern.ch/record/265610/files/P00024340.pdf}}.

\bibitem[P47]{P47}
B. Pontecorvo, 
Nuclear capture of mesons and the meson decay,
{\it Phys. Rev.} {\bf 72}:3 (1947) 246-247.

\bibitem[P57]{P57}
H. Postma et al.,
Asymmetry of the positon emission by polarized $^{58}$Co-nuclei,
{\it Physica} {\bf 23} (1957) 159-160, received on February 25, 1957.

\bibitem[Re]{Re}
E. Recami,
Majorana, the Neutron, and the Neutrino: Some elementary historical remarks,
{\it Hadronic J.} {\bf 40} (2017) 149-185,
{\tt arXiv:1712.02209[physics.gen-ph]}.

\bibitem[RR53]{RR53}
B.M. Rustad, S.L. Ruby,
Correlation between electron and recoil nucleus in He$^6$ decay,
{\it Phys. Rev.} {\bf 89}:4 (1953) 880-881.

\bibitem[RR55]{RR55}
B.M. Rustad, S.L. Ruby,
Gamow-Teller interaction in the decay of He$^6$,
{\it Phys. Rev.} {\bf 97}:4 (1955) 991-1002.

\bibitem[S57]{S57}
J.J. Sakurai,
Mass reversal and weak interactions,
{\em Nuovo Cimento} {\bf 7}:5 (1958) 649-660,
received on October 31, 1957.

\bibitem[Salam57]{Salam57}
A. Salam,
On parity conservation and neutrino mass,
{\it Il Nuovo Cimento Ser.~X} {\bf 5} (1957) 299-301,
received on November 15, 1956.

\bibitem[SW]{SW}
R.F. Streater, A.S. Wightman,
\textit{PCT, Spin and Statistics, and All That},
Princeton Univ. Press, 2000.

\bibitem[SM1]{SM1}
E.C.G. Sudarshan, R.E. Marshak,
The nature of the four-fermion interaction, {\bf in}:
Proceedings of the Padua-Venice Conference on Mesons and Newly Discovered Particles, 22-27 September 1957, N. Zanichelli (ed.), Bologna 1958, p. V-14,
%Suppl. Nuovo Cimento (to be published)
%(Societ\`a Italiana di Fisica, Padova-Venezia, 1958),
available e.g. at
\href{https://web2.ph.utexas.edu/~gsudama/pub/1958_005.pdf}{\footnotesize{{\tt https://web2.ph.utexas.edu/$\sim$gsudama/pub/1958\_005.pdf}}};
\noindent
reprinted {\bf in}: P.K. Kabir, {\em The Development of Weak Interaction Theory},
Gordon and Breach, New York 1963, pp. 118-128.

\bibitem[SM2]{SM2}
E.C.G. Sudarshan, R.E. Marshak,
Chirality invariance and the universal Fermi interaction, %Letter to the Editor,
{\it Phys. Rev.} {\bf 109}:5 (1958) 1860-1862.

\bibitem[SM3]{SM3}
E.C.G. Sudarshan, R.E. Marshak,
Origin of the Universal V-A Theory,
{\bf in}: Proceedings of the Wingspread Conference
"50 Years of Weak Interactions", University of Wisconsin, Madison %, Wisconsin
(1984), pp. 1-15;
and {\bf in}: AIP Conference Proceedings 300 "Discovery of Weak Neutral Currents:
the Weak Interaction Before and After", A.K. Mann, D.B. Cline (eds.),
AIP, New York (1994), pp. 110-124.

\bibitem[SM4]{SM4}
E.C.G. Sudarshan, R.E. Marshak, Conserved Currents in Weak Interactions,
Frontiers of Physics (Proc. of The Landau Memorial Conference, Tel Aviv, Israel, 6-10 June 1988),
E. Gotsman, Y. Ne'eman, A. Voronel (eds.),
Pergamon Press, Oxford (1990), pp. 169-182.

\bibitem[Th1894]{Th1894}
Sir William Thomson Lord Kelvin,
{\em The Molecular Tactics of a Crystal}, Clarendon Press (1894).

\bibitem[T55]{T55}
J. Tiomno,
Mass reversal and the universal interaction,
{\it Nuovo Cimento} {\bf 1}:1 (1955) 226-232.

\bibitem[TW49]{TW49}
J. Tiomno, J.A. Wheeler,
Energy spectrum of electrons from meson decay,
{\it Rev. Mod. Phys.} {\bf 21}:1 (1949) 144-152.

\bibitem[T11]{T11}
I. Todorov,
Clifford algebras and spinors,
{\it Bulg. J. Phys.} {\bf 38}:1 (2011) 3-28;
{\tt arXiv:1106.3197 [math-ph]}.


\bibitem[W09]{W09}
S. Weinberg,
V-A was the key,
{\it J. Phys.: Conf. Ser.} {\bf 196} (2009) 012002.

\bibitem[W29]{W29}
H. Weyl,
Gravitation and the electron,
{\it Proc. Natl. Acad. Sci. USA} {\bf 15}:4 (1929) 323-334;
H. Weyl,
Elektron und Gravitation. I. (in German),
{\it Z. Physik} {\bf 56} (1929) 330-352.

\bibitem[WWW52]{WWW52}
G.C. Wick, A.S. Wightman, E.P. Wigner,
The intrinsic parity of elementary particles,
{\it Phys. Rev.} {\bf 88} (1952) 101-105.

\bibitem[W27]{W27}
E.P. Wigner, Einige Folgerungen aus der Schr\"odinger-schen Theorie f\"ur die Termstrukturen,
{\em Z. Physik} {\bf 43} (1927) 624-652.

\bibitem[Wu57]{Wu57}
C.S. Wu, E. Ambler, R.W. Hayward, D.D. Hoppes, R.P. Hudson,
Experimental test of parity conservation in beta decay,
\textit{Phys. Rev.} \textbf{105}:4 (1957) 1413-1415.

\bibitem[Y82]{Y82}
C.N. Yang,
The discrete symmetries P, T and C,
{\it J. de Physique Colloques} {\bf 43} (C8) (1982) 439-451.

\bibitem[Y35]{Y35}
H. Yukawa,
On the Interaction of Elementary Particles,
{\it Proc. Phys. Math. Soc. Jap.} {\bf 17} (48) (1935) 48-57;
{\it Prog. Theor. Phys. Suppl.} {\bf 1} (1955) 1-10.

\end{thebibliography}
\end{document}